\def\jref#1 #2 #3 #4 {{\par\noindent \hangindent=3em \hangafter=1
    #1, {\it#2}, {\bf#3}, #4.}}
\def\jreftitle#1 #2 #3 #4 #5 {{\par\noindent \hangindent=3em \hangafter=1
    #1, {``#2,''} {\it#3}, {\bf#4}, #5.}}
\def\ref#1{{\par\noindent \hangindent=3em \hangafter=1 #1.}}
\def\references{\subsection*{REFERENCES}
\bgroup\parindent=0pt\parskip=\itemsep
\def\refpar{\par\hangindent=1.2em\hangafter=1}}
\def\endreferences{\refpar\egroup}

\def\@biblabel#1{\relax}
\def\@cite#1#2{#1\if@tempswa , #2\fi}
\def\reference{\relax\refpar}
\def\@citex[#1]#2{\if@filesw\immediate\write\@auxout{\string\citation{#2}}\fi
\def\@citea{}\@cite{\@for\@citeb:=#2\do
{\@citea\def\@citea{,\penalty\@m\ }\@ifundefined
{b@\@citeb}{\@warning
{Citation `\@citeb' on page \thepage \space undefined}}%
{\csname b@\@citeb\endcsname}}}{#1}}
\def\aj{\ref@jnl{AJ}}
\def\araa{\ref@jnl{ARA\&A}}
\def\apj{\ref@jnl{ApJ}}
\def\apjl{\ref@jnl{ApJ}}
\def\apjs{\ref@jnl{ApJS}}
\def\applopt{\ref@jnl{Appl.Optics}}
\def\apss{\ref@jnl{Ap\&SS}}
\def\aap{\ref@jnl{A\&A}}
\def\aapr{\ref@jnl{A\&A~Rev.}}
\def\aaps{\ref@jnl{A\&AS}}
\def\azh{\ref@jnl{AZh}}
\def\baas{\ref@jnl{BAAS}}
\def\jrasc{\ref@jnl{JRASC}}
\def\memras{\ref@jnl{MmRAS}}
\def\mnras{\ref@jnl{MNRAS}}
\def\pra{\ref@jnl{Phys.Rev.A}}
\def\prb{\ref@jnl{Phys.Rev.B}}
\def\prc{\ref@jnl{Phys.Rev.C}}
\def\prd{\ref@jnl{Phys.Rev.D}}
\def\prl{\ref@jnl{Phys.Rev.Lett}}
\def\pasp{\ref@jnl{PASP}}
\def\pasj{\ref@jnl{PASJ}}
\def\qjras{\ref@jnl{QJRAS}}
\def\skytel{\ref@jnl{S\&T}}
\def\solphys{\ref@jnl{Solar~Phys.}}
\def\sovast{\ref@jnl{Soviet~Ast.}}
\def\ssr{\ref@jnl{Space~Sci.Rev.}}
\def\zap{\ref@jnl{ZAp}}
\def\lefevre{Le\thinspace F\`evre~}
\def\ms0440{MS\thinspace 0440$+$0204}
\def\etal{\it et al.~\rm}
\def\solar{\footnotesize \sun}

\def\deg{\hbox{$^\circ$}}
\def\sun{\hbox{$\odot$}}

\def\la{\mathrel{\hbox{\rlap{\hbox{\lower4pt\hbox{$\sim$}}}\hbox{$<$}}}}
\def\ga{\mathrel{\hbox{\rlap{\hbox{\lower4pt\hbox{$\sim$}}}\hbox{$>$}}}}

\def\arcmin{\hbox{$^\prime$}}
\def\arcsec{\hbox{$^{\prime\prime}$}}

\documentstyle[11pt]{article}
\begin{document}
\begin{center}
{\Large\bf The EMSS catalog of X-ray-selected clusters of galaxies. I.\\
An atlas of CCD images of 41 distant clusters}\\
\vskip 20pt
{\Large {\sc I. M. Gioia}\footnote[1]{
On leave from Istituto di Radioastronomia del CNR,
Via Gobetti 101, I-40129, Bologna, ITALY}$^,$\footnote[2]{and
the Harvard-Smithsonian Center for Astrophysics, 60 Garden Street,
Cambridge, MA USA 02138}
and {\sc G. A. Luppino
}}
\vskip 10pt
{\large Institute for Astronomy, University of Hawaii\\
2680 Woodlawn Drive, Honolulu, Hawaii 96822\\
Internet: gioia@ifa.hawaii.edu, ger@ifa.hawaii.edu}
\end{center}
\vskip 20pt

\begin{center}
{\large Submitted to the {\large\it Astrophysical Journal Supplement
Series} \\
\vskip 20pt
Received: \underline{December 1993}; ~~Accepted: \underline{February 1994}
}
\end{center}

\begin{center}
{\large\bf Abstract\vspace{-.5em}\vspace{0pt}}
\end{center}

{\quotation
An atlas of deep, wide-field $R$-band CCD images of a
complete sample of distant, X-ray-selected clusters of galaxies is presented.
These clusters are the 41 most distant ($z\geq 0.15$) and most
X-ray luminous ($L_x \geq 2\times 10^{44}$ erg\thinspace s$^{-1}$) clusters
in the {\it Einstein Observatory} Extended Medium Sensitivity
Survey (EMSS) catalog  that are observable from Mauna Kea
($\delta > -40^{\rm o}$).  The sample spans a redshift range of
$0.15 \leq z \leq 0.81$ and includes at least 2 and possibly as many as
6 rich clusters with $z>0.5$. For the most part, the data are of
superior quality, with a median seeing of $0''.8$ FWHM and coverage of
at least 1~Mpc~$\times$~1~Mpc in the cluster frame ($H_0=50$; $q_0=1/2$).

In addition, we update the available optical, X-ray and radio data on
the entire EMSS sample of 104 clusters.
We outline the cluster selection criteria in detail, and emphasize
that X-ray-selected cluster samples may prove to be more useful for
cosmological studies than optically selected samples.
The EMSS cluster sample in particular can be
exploited for diverse cosmological investigations, as demonstrated
by the detection of evolution in the X-ray luminosity function
previously reported, and more recently by the discovery of a
large number of gravitationally lensed images in these clusters.

\par
}
\clearpage
\begin{center}
{\Large\bf 1.~Introduction\vspace{-.5em}\vspace{0pt}}
\end{center}
\vskip 10pt
Distant clusters of galaxies provide information about the early
universe. They can be used for cosmological tests such as those based
on the magnitude-redshift relation or the Sunyaev-Zeldovich effect.
They can be used as markers of large-scale structure. Changes in their
properties as a function of look-back time are a record of cosmic
evolution. In the past, cluster samples have been selected
from optical surveys such as those of Abell (1958),
Abell, Corwin \& Olowin (1989), Gunn, Hoessel \& Oke (1986), or
Couch \etal (1991).  However, optical selection involves the
detection of a surface density enhancement of galaxies against
a rich and variable background, which sometimes leads to false
identifications due to superposition effects (see Frenk \etal
1990), especially at high redshifts.

Superposition effects can be avoided if clusters are selected
based on the X-ray emission from the intracluster gas,
rather than on the number density of optical galaxies.
Clusters of galaxies are known to be very luminous X-ray sources,
with the X-ray luminosities often exceeding $3 \times
10^{43}$ erg s$^{-1}$. Projection effects are much less severe in the
X-ray than in the visible.  First, no cluster will be lost due to
superposition onto a smaller than average background because the
surface brightness of the central cluster X-ray emission is several hundred
times that of the X-ray background. Second, few clusters will be
found which are actually the superposition of
less rich systems because the X-ray luminosity of a cluster scales
approximately as the cube of the richness (Bahcall 1977).
Moreover, the X-ray temperature, which is analogous to the galaxy velocity
dispersion, will be unaffected by projection.  X-ray surveys and
source detection methods are automated so that selection effects are well
defined and quantifiable.

In 1992, we began an observational program to obtain
deep, wide-field  $B$ and $R$ CCD images of the most X-ray luminous,
distant clusters selected from the {\it Einstein Observatory}
Extended Medium Sensitivity Survey (EMSS) catalog
(Gioia \etal 1990a; Stocke \etal 1991).
Until the ROSAT
North Ecliptic Pole region of the all-sky survey is completely
identified, the EMSS is the only large (835 sources)
and sensitive ($F_{x} \geq 7 \times 10^{-14}$ erg cm$^{-2}$ s$^{-1}$
in the 0.3$-$3.5 keV band) X-ray survey from which a sample of distant
clusters can be extracted.
The goals of our survey were two-fold.  Our primary motivation was to
search for gravitationally lensed arcs in a sample of X-ray selected
clusters. We also intended to compile a catalog of optical
data on a complete sample of rich distant clusters that were not
selected by traditional optical methods. The first results
from the arc survey have already appeared in the literature  (Luppino
and Gioia 1992; Luppino \etal 1993; \lefevre \etal 1994) or are in
preparation (Luppino \etal 1994). Here we present for the first
time the deep CCD images acquired by us for the 41 clusters that
constitute the arc survey sample. Subsequent papers will investigate
the optical properties of the clusters that will be
derived from the analysis of the arc survey data.

In Section 2 we present and discuss the entire EMSS sample of 104
clusters from which the 41-cluster arc survey sample is extracted.
Drawing upon new ROSAT data and additional imaging and spectroscopic
data, we have updated the EMSS cluster list of 104 clusters
(Gioia \etal 1990a; Stocke \etal 1991; Henry \etal 1992). In this
section, we also update the status of the remaining EMSS unidentified
fields that might possibly enhance the cluster sample. Our optical
observations of the 41 most luminous
and distant EMSS clusters, and a description of the data reduction are
given in Section 3. In Section 4 we discuss qualitative properties
of our X-ray selected sample. Quantitative properties will be
evaluated in subsequent papers in this series. Throughout, $H_{0}$=50 km
s$^{-1}$ Mpc$^{-1}$ and $q_0=1/2$ are assumed.

\clearpage
\begin{center}
{\Large\bf 2.~The Sample\vspace{-.5em}\vspace{0pt}}
\end{center}
\vskip 10pt
\par\noindent
{\Large\it 2.1~The EMSS galaxy cluster sample}
\vskip 6pt
The 41 clusters whose images appear in this atlas
are extracted from the EMSS. Briefly,
the EMSS is a flux limited and homogeneous sample
containing 835 sources discovered
serendipitously in Einstein IPC images at high Galactic latitude. The limiting
sensitivities of the IPC fields range from 5 $\times 10^{-14}$
to 3 $\times 10^{-12}$ erg cm$^{-2}$ s$^{-1}$ in the 0.3 $-$ 3.5 keV
band. Details of the X-ray survey can be found in Gioia \etal (1990a).
The optical identifications have been discussed in Stocke \etal (1991).
Only 25 sources are still optically unidentified. It
is thus possible to extract samples of X-ray-selected
objects exclusively defined by their X-ray properties and
suitable for statistical studies. There is a total of 104
clusters of galaxies in the EMSS. This count is corrected in line
with recently acquired information on five
so-called ``cooling flow galaxies''. This term was
adopted by Stocke \etal (1991) to indicate objects with
low-ionization, optical emission line spectra very similar
to those of central galaxies in cooling flow clusters, but
with no surrounding rich cluster present. Additional optical data and
new ROSAT HRI pointed observations have helped us to  clarify the
nature of these sources. Three of them are now tentatively classified as
clusters due to the presence of galaxies around them (MS1125.3+4324,
MS1209.0+3917, MS1317.0-2111). The remaining two, MS1826.5+7256 and
MS1019.0+5139, are identified with a flaring M star and a peculiar type
of AGN or a possible BL Lac object respectively (Stocke, private
communication) and will be described elsewhere.

When selecting in X-rays there are a number of effects in the data
that must be corrected. These effects include absorption by the Milky
Way, the different sky coverage for different flux limits, correction
for lost flux due to the finite $2.4\thinspace \arcmin \times 2.4
\thinspace \arcmin$ detection cell, and any variation of these effects
with the different redshifts of the sources. We describe each of these
corrections in turn. We recall here that the EMSS uses the so
called M-DETECT algorithm to find sources. In this method the
background is computed from a global map of the detector, hence
sources are not lost because their extended flux distribution
increases the apparent background around them (see Gioia \etal
1990a for details). The flux for each source was
converted from IPC counting rates with a Raymond-Smith
thermal spectrum of 6 keV and  corrected for
absorption using the neutral hydrogen values from the survey of Stark
\etal (1984). Most of the sky was observed through a small range of
N$_{H}$, resulting in a negligible bias (see Maccacaro \etal 1988
and Zamorani \etal 1988). K corrections were small
for this sample and were calculated assuming for simplicity a power law
spectrum with energy index of 0.5, which approximates a 6 keV thermal
spectrum in the 0.3$-$3.5 keV band. The largest correction accounts for
the finite size of the detection cell. This correction varies with redshift and
must be included in any calculation involving the flux of the
sources. The EMSS sample of clusters is in fact a surface-brightness
limited sample rather than a flux-limited sample. Thus what we and
other authors who have used the cluster sample (see Gioia \etal 1990b;
Donahue, Stocke \& Gioia 1992; Henry \etal 1992) have
attempted to do is to take into account the spatial variations
of the X-ray surface brightness in order to correct the
surface brightness limited sample to a flux limited sample.
To do this, an estimate of the spatial distribution
of X-ray flux for distant clusters is derived based upon the
observed structure of nearby clusters. Since cluster X-ray emission is
extended and comparable to the size of the detection cell, this
correction is significant at low redshift and becomes less important,
but not negligible, at high redshift (see values
for detection cell flux and corrected total flux in
columns (5) and (6) of Table~1). Gioia \etal (1990b) and Henry
\etal (1992) have described in detail the prescription for correcting
this selection effect (see also the discussion in Donahue, Stocke \&
Gioia 1992). Briefly, a $\beta$ = 2/3 model for cluster surface brightness
distribution is adopted. The surface brightness is then integrated
over the square  $2.4\thinspace \arcmin \times 2.4 \thinspace \arcmin$
detection cell to calculate the fraction of total flux deposited in
the detection cell. A core radius, observationally determined from
a large number of clusters observed by non-imaging, large beam
experiments and which have IPC imaging data, is then used to
correct the observed luminosity to the total ``true'' luminosity for
the flux lost outside the detection cell.
The luminosities in column (7) of Table~1 have been computed adopting
this correction.

Another point worth mentioning is the presence of
cooling flow clusters in the sample. It has been suggested (Pesce
\etal 1990; Edge \etal 1992) that most of the clusters
found in the EMSS contain cooling flows given the point-source
detection algorithm used in the survey. Donahue, Stocke \& Gioia (1992)
have shown that some non-cooling flow
EMSS clusters have small core radii, thus implying that the sample
is not necessarily dominated by cooling flow clusters.
Besides, Donahue, Stocke \& Gioia (1992) suggest that many clusters
with large core radii may have
low luminosities, and thus would not be detected in X-ray imaging
surveys even if they were more compact. More compelling  evidence
that the EMSS does not miss clusters which are not cooling flow clusters
is given by the agreement between the X-ray luminosity function of the
EMSS clusters with redshifts between 0.14 and 0.2, and the luminosity
function derived by earlier studies (Piccinotti \etal
1982) which use large-beam, non-imaging detector fluxes.
The agreement seems to indicate that this bias is not
present in our sample, or at least it is at work in the same manner
in the non-imaging data (see also Henry \etal 1992 and Donahue,
Stocke \& Gioia 1992 for an extensive discussion of this last point).

The X-ray emission of clusters possessing enough counts to allow
a meaningful analysis of their morphologies ranges from smooth
extended features to more complex structures with several
subcondensations (X-ray iso-contour maps for sources with more
than 100 counts are published in Gioia \etal 1984, Gioia \etal 1987
and Gioia \etal 1990a). The X-ray emission is often centered on
a bright, optically dominant galaxy. However, there are also several
cases where there are two approximately equally bright galaxies and
the cluster appears to be irregular, and many cases where no dominant galaxy
is visible (see CCD images in this paper) and the cluster is not rich.
In fact both rich clusters and poor groups of galaxies, with
or without cD galaxies, seem to be selected in X-rays.
Optical selection is based on high apparent galaxy concentration
and thus would select only rich clusters in the absence of confusion.

The resulting sample of 104 clusters is given in Table~1
with the above described corrections applied. Columns list the observed
and derived parameters for each cluster as follows:
\begin{enumerate}
\item EMSS Name. Those clusters preceded by an asterisk make up the 41
cluster arc survey sample whose images are presented in Figures 2.1
through 2.41.
\item Other Name.
\item Right Ascension and Declination in 1950.0 coordinates of
the cluster optical position. Most of these positions were measured
on the POSS or SRC J plates (Stocke \etal 1991) and are accurate
to $\pm 5\arcsec$. When the brightest cluster member has been
detected in radio, then the VLA position is given accurate
to $\pm 1\arcsec$.
\item Redshift of the cluster ($\pm 0.003$). Those redshifts enclosed
by parenthesis are tentative values.
\item X-ray flux, in units of $10^{-13}$
erg cm$^{-2}$ s$^{-1}$ (0.3$-$3.5 keV) in the
$2.4\thinspace \arcmin \times 2.4\thinspace \arcmin$ detection cell
with correction only for vignetting and mirror scattering.
\item Corrected X-ray flux (same units).
\item Corrected X-ray luminosity in units of $10^{44}$ erg s$^{-1}$
(0.3$-$3.5 keV).
\item Radio flux at 6 cm or a 5 $\sigma$ upper limit.
\end{enumerate}
Radio observations at 6 cm with the VLA in a variety of telescope
configurations were performed for the 98 clusters in the declination range
$\delta \geq -43 \deg$. A detection
rate of 50\% is reached with  most of the radio detections unresolved
and coincident with the brightest galaxy in the cluster. There are,
however, a few
resolved structures which have been published in Gioia \etal (1987).
Table 2 presents the radio position and 6 cm flux for all the
radio sources detected in the cluster fields and their association
with the cluster galaxies as indicated in the Note column.

\vskip 10pt
\par\noindent
{\Large\it 2.2~The EMSS distant cluster sample for the arc survey}
\vskip 6pt
The 41 cluster subsample chosen for the gravitationally lensed
arc survey (Luppino \etal 1993, 1994), whose
optical images are presented here, is subject to the following
additional restrictions. First, the sources must have a declination
$\delta \geq -40 \deg$ to be observable from Mauna Kea. Second,
the fluxes of the sources in the
$2.4\thinspace \arcmin \times 2.4 \thinspace \arcmin$ detection cell
(flux in column (5) of Table 1) must be $F_x^{Det}\geq 1.33 \times 10^{-13}$
erg cm$^{-2}$ s$^{-1}$ after converting from IPC counting rates with
a thermal spectrum of 6 keV temperature and correcting for
the galactic absorption in
the direction of each source, but with no IPC point response function
correction applied. These two selection criteria restrict the number of
unidentified sources to only 9 out of 733 sources, an identification
rate of 98.8 \%. Third, the redshift of the cluster must be
$z\geq 0.15$ to insure a distant sample for which at least
1 Mpc $\times$ 1 Mpc would be imaged onto our
$7.5\thinspace \arcmin \times 7.5\thinspace \arcmin$ CCD field.
Fourth, the X-ray luminosity must be
greater than $L_{x} = 2 \times 10^{44}$ erg s$^{-1}$ to select for
deep potential wells which are most likely to exhibit gravitational lensing.
These selection criteria result in a
subsample of 41 clusters which are identified with an asterisk in Table~1.

\vskip 10pt
\par\noindent
{\Large\it 2.3~The remaining EMSS unidentified sources}
\vskip 6pt

The nine still unidentified sources in the restricted EMSS
catalog of 733 sources are given in Table~3.

\begin{center}
\samepage
{\normalsize Table 3. Unidentified Sources}

\small
{\sf
\begin{tabular}{cc}
\hline\hline
\multicolumn{1}{c}{(1)} &
\multicolumn{1}{c}{(2)} \\
\multicolumn{1}{c}{EMSS Name} & \multicolumn{1}{c}{$F_x^{Det}$ \scriptsize
$\times 10^{-13}$} \\
\hline
 & \\
MS0235.6$+$1631& 1.74 \\
MS0354.2$-$3658& 3.30 \\
MS0501.0$-$2237& 3.19 \\
MS1237.9$-$2927& 4.86 \\
MS1411.0$-$0310& 1.83 \\
MS2136.1$-$1509& 2.50 \\
MS2144.2$+$0358& 1.74 \\
MS2223.8$-$0503& 1.59 \\
MS2225.7$-$2100& 1.33 \\
\hline
\end{tabular}
}
\end{center}

\noindent
This list updates the one that appeared as Table~1 in Henry \etal
(1992). One of the previously unidentified sources, MS1610.4$+$6616,
has been classified as a distant cluster and is included in
our sample. Two other unidentified sources have been associated with
active galactic nuclei (Maccacaro \& Wolter, private communication).
No plausible optical counterparts have been found in the IPC error
circles of the nine sources in Table~3 despite optical observations of the
candidates visible on the POSS, or on moderate resolution CCD images
that we have acquired during the identification process of the 835
EMSS sources.
The IPC has only a positional accuracy of $\sim$ 50$\arcsec$
(90\% confidence radius) so we have obtained ROSAT PSPC observations for
seven of the unidentified sources. Reduction of the ROSAT data is still under
way, but from an inspection of the new X-ray data there are no
more than two sources which could be identified with clusters
of galaxies (Gioia \etal 1994). Deeper imaging and/or
spectroscopy is necessary to identify unambiguously these remaining
nine sources. The small number of
unidentified sources which could be associated with galaxy clusters
makes the EMSS cluster sample a well defined, homogeneous
and statistically reliable list of clusters which can be
used for cosmological studies.

\begin{center}
{\Large\bf 3.~Observations and Data Reduction\vspace{-.5em}\vspace{0pt}}
\end{center}
\vskip 10pt

The optical observations were carried out during the period from May 1992 to
November 1993 using a thinned Tektronix $2048 \times 2048$ pixel CCD
mounted at the f/10 focus of the Univerity of Hawaii 88-inch
telescope on Mauna Kea.
The image scale was $0''.22$/pixel and the resulting field of view
was $7.5\thinspace \arcmin \times 7.5\thinspace \arcmin$. Images
were taken through (Johnson) $B$ and (Kron-Cousins) $R$ filters.
The data were calibrated using the photometric standards of Landolt
(1992). Exposures ranged from tens of minutes for the relatively
bright, nearby
clusters to several hours for the most distant clusters. All the data
were taken with the same instrument configuration---telescope, filters
and detector. For the most part, these data are of superior quality.
The median seeing for the entire 41 cluster sample is $0''.8$ FWHM
(see Figure 1).
\begin{figure*}[t]
\vskip 6.0truein
\hspace{-1.0truein}
\vspace{-1.4truein}
\includegraphics{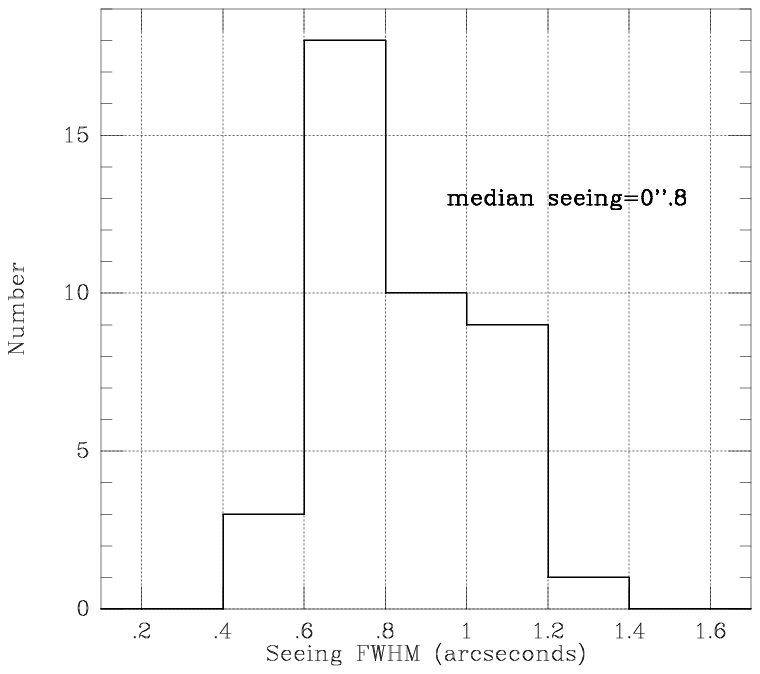}
\par\noindent
{\small{\bf Figure 1.}---Histogram of the $R$-band seeing
for the 41 EMSS cluster images shown in Figure 2. The mean seeing
is $0''.88$ FWHM and the median seeing is $0''.8$ FWHM.
}
\end{figure*}

In order to build up our long exposures, we took a number of short
exposures  (typically 600\thinspace s or 900\thinspace s each) with
the telescope shifted between exposures, allowing us to assemble a
median-filtered stack of the disregistered images to use for
flattening the data. These multiple exposures were then shifted
into registration and combined while being cleaned of cosmic rays.

We chose the $B$ and $R$ bandpasses for the following reasons.
The $R$ bandpass is ideal for observing the distant clusters
in the redshift range spanned by the EMSS sample.  The center
wavelength of the $R$ filter occurs at the peak of the CCD quantum
efficiency and the night sky is darker than a redder bandpass such as $I$.
Although gravitationally lensed arcs are known to
be relatively blue, they are still more easily detected in
the $R$ band since their average $B-R$ color is roughly 1.3 (Soucail 1992).
The additional $B$ band image is taken to recognize
any gravitational arcs by their relative blue color compared to the
red cluster galaxies.

\begin{center}
{\Large\bf 4.~Results and Discussion\vspace{-.5em}\vspace{0pt}}
\end{center}
\vskip 10pt

Figures 2.1 through 2.41 show the resulting stacked CCD images
obtained for the arc survey clusters.
Additional parameters are given in each figure caption
such as size and scale of the image, field of view at the redshift
of the cluster (both in angular and linear units) plus any other
interesting observational parameter (exposure time, seeing, etc.).
For most of the clusters, we have extracted 1024 $\times$ 1024
or 1200 $\times$ 1200 subarrays centered on the cluster.  These
subarray images allow one to see at least the central 1 Mpc of
the cluster.  The coordinate grid surrounding each image is
measured in arcseconds with the origin centered on the optical position given
in Table~1.  In all cases North is up and East is to the left as
indicated at the upper left-hand part of the image.

A close inspection of the CCD images shows that the optical richness
of X-ray selected clusters are very diverse, ranging from poor
groups of galaxies (e.g. MS0043.3$-$2531, MS0433.9$+$0957) both with
and without an optically dominant galaxy, to
``extremely rich'' clusters with hundreds of members (e.g. MS0015.9$+$1609,
MS0451.6$-$0305, MS1008.1$-$1224, MS1054.4$-$0321, MS1358.4$+$6245).
Different morphologies are also present: ``binary'' clusters with
two central galaxies of comparable brightness (e.g. see MS1231.3$+$1542),
clusters with ``linear'' structures of galaxies
as in the case of MS0015.9$+$1609 and MS1054.4$-$0321, loose, irregular
(but nevertheless rich) clusters like MS1241.5$+$1710, or compact clusters
(MS1201.5$+$2824, MS1208.7$+$3928). This classification is only subjective
and qualitative at this time. Quantitative properties will be evaluated and
presented in subsequent papers in this series.

Twenty Abell and eleven Zwicky clusters were ``rediscovered'' in the EMSS.
An EMSS selection criterion causes the exclusion of
Abell clusters with a distance class D $\leq 3$ from our sample,
since most of the nearby clusters were chosen as targets of IPC
observations and consequently were excluded from the survey of
serendipitous X-ray sources.
During this survey a new, distant cluster was added to the list,
MS1610.4$+$6616. This object was previously one of the unidentified
sources. Even if the spectroscopic data in hand are
not satisfactory for a firm redshift determination, the cluster is
undoubtedly a distant one as can be clearly seen from the CCD image in
Figure 2.33. On the other hand there are two objects MS1209.0$+$3917
and MS1333.3$+$1725 whose identification with a cluster is
questionable. The first source was identified as a ``cooling flow
galaxy'' in Stocke \etal (1991) and is here presented as a
cluster due to the presence of a few galaxies around it.
Both clusters look like very poor systems (unless they are
very distant ones). We would like to stress that these two
identifications are tentative and the need for more spectroscopy
or deeper imaging is in order.

The high quality of the CCD images will allow us to study the
brightest cluster galaxies. In several cases (MS0440.5$+$0204,
MS1244.2$+$7114, MS2124.7$-$2206) the central galaxy is
composed of multiple nuclei embedded in a common
envelope. In some cases it consists of double nuclei (e.g. MS0906.5$+$1110,
MS1910.5$+$6736) or triple nuclei (e.g. MS0353.6$-$3642,
MS1426.4$+$0158), in other cases the central galaxy is highly
structured like in MS1512.4$+$3647. The multiple-nuclei systems
(e.g. MS1244.2$+$7114 and MS0440.5$+$0204) might be
examples of cD galaxies under construction, as pointed out by Luppino
\etal (1993) for MS0440.5$+$0204, where the central galaxy may be
cannibalizing smaller cluster members. Annis (1994) has pointed out an
evolutionary effect in the morphologies of the brightest cluster
members of the EMSS (z$=$1/3 sample) with respect to the z$=$0 HEAO-1 A2
cluster sample (Piccinotti \etal 1982). The double morphologies are
more often seen in the EMSS than in the present epoch (z$=$0) HEAO$-$1
brightest cluster galaxies, the triple systems seen in the
EMSS are non-existent in the HEAO$-$1 sample. This fact
could be taken as a preliminary evidence of morphological evolution
in the brightest cluster galaxies
at larger distances (z$\geq$0.3).

\begin{figure*}[t]
\vskip 5.2truein
\hspace{-1.0truein}
\vspace{-1.4truein}
\includegraphics{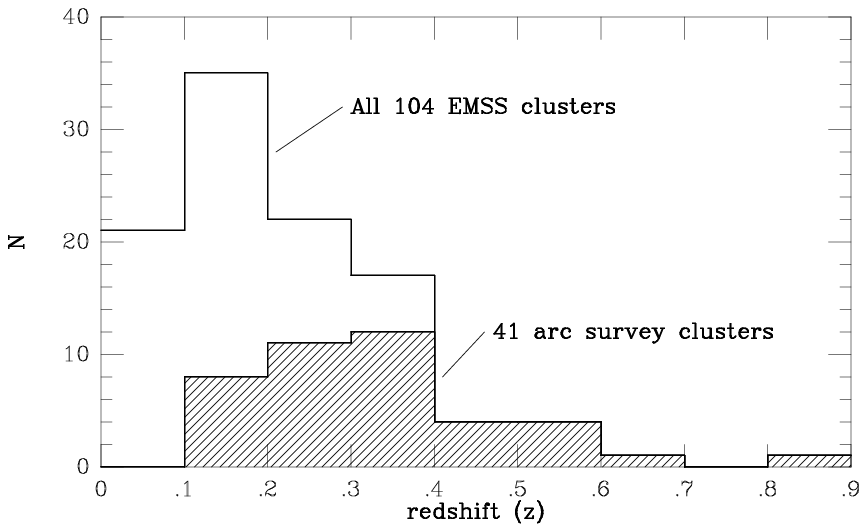}
\par\noindent
{\small{\bf Figure 3.}---Redshift histogram for the entire EMSS
cluster sample and for the 41 arc survey clusters.
The mean redshift is $<z>=$0.22 and $<z>=$0.32 respectively.

}
\end{figure*}

As can be seen in the redshift histogram of Figure 3, the EMSS
clusters constitute an excellent sample for cosmological studies.
The average redshift
for the full sample is $<z> =$ 0.22 with a dispersion
$\sigma =$ 0.14. The arc survey sample consists of the most luminous
and distant of these clusters, and we note that all of the EMSS clusters
with $z>0.4$ meet the selection criteria for the arc survey sample.
The redshift range encompassed by the
clusters (possibly beyond $z$ of 0.8) will allow the use of the sample
for several projects to study the properties of  clusters
of galaxies at intermediate to large redshifts.
Distant clusters provide important constraints on cosmological
models. This same EMSS distant cluster sample provided evidence for
the evolution of the X-ray luminosity function (Gioia \etal 1990b,
Henry \etal 1982).
The nature of the observed evolution (fewer high X-ray luminosity
clusters in the past) is consistent with hierarchical
structure formation. Henry  \etal (1992) used the results
to determine the shape and normalization of the mass fluctuation
spectrum.

One of the motivations for the imaging survey was the search for
gravitationally-lensed arcs. We already mentioned the major
improvement of an X-ray selected sample compared to lens surveys based
on optically selected clusters. In addition, the X-ray selection
should have the property that the same kinds of objects are selected
at differing redshifts. To draw inferences about clusters from
a study of gravitational lensing, one has to understand
whether the individual clusters are representative of the class
as a whole. The X-ray selection assures the presence
of a deep potential well, and should therefore
provide this representative sample.
The fairly large number of lensed images that we are finding in
this survey (Luppino \etal 1993, 1994; \lefevre \etal 1994) gives us
confidence that the EMSS cluster sample can be widely exploited
for cosmological investigations.

\begin{center}
{\Large\bf Acknowledgements\vspace{-.5em}\vspace{0pt}}
\end{center}
\vskip 10pt
We thank Jim Annis for assisting with some of the observing.
This work was partially supported by NASA grant NAG5-1880 and by NSF
grant AST-9119216. The UH CCD cameras were fabricated with
NSF grant AST-9020680. Radio data presented in this paper
were obtained with the NRAO Very Large Array. The National Radio
Astronomy Observatory (NRAO) is operated by Associated Universities,
Inc., under contract with the National Science Foundation (NSF).
\onecolumn
\pagestyle{empty}
\begin{center}
\samepage
{\normalsize Table 1.  EMSS X-ray-selected galaxy cluster sample }
\footnotesize
{\sf
\begin{tabular}{rcrrrrrrcrrrr}
\hline\hline
\multicolumn{1}{c}{(1)}&
(2)&
\multicolumn{6}{c}{(3)}&
(4)&
\multicolumn{1}{c}{(5)}& \multicolumn{1}{c}{(6)} & \multicolumn{1}{c}{(7)} &
\multicolumn{1}{c}{(8)} \\
& & & & & & & & & \multicolumn{3}{c}{X-ray Data} & \multicolumn{1}{c}{$F_R$} \\
&\multicolumn{1}{c}{Other} & \multicolumn{6}{c}{Optical Coordinates (1950.0)} &
&
\multicolumn{1}{c}{$F_x^{Det}$} & \multicolumn{1}{c}{$F_x^{Tot}$} &
\multicolumn{1}{c}{$L_{x}$} & \multicolumn{1}{c}{(6\thinspace cm)} \\
\multicolumn{1}{c}{EMSS Name} & \multicolumn{1}{c}{Name} &
\multicolumn{3}{c}{R.A.} & \multicolumn{3}{c}{Dec.} &
\multicolumn{1}{c}{$z$} &
\multicolumn{1}{c}{\scriptsize $\times 10^{-13}$} &
\multicolumn{1}{c}{\scriptsize $\times 10^{-13}$} &
\multicolumn{1}{c}{\scriptsize $\times 10^{44}$} & \multicolumn{1}{c}{mJy} \\
\hline
& & & & & & & & & & & & \\
 MS0002.8$+$1556  &           &00 &02 &49.9  &$+$15 &56 &24.3   &.116  & 7.28
& 28.47  & 1.643  & $<$0.9   \\
 MS0007.2$-$3532  &           &00 &07 &14.5  &$-$35 &33 &11.8   &.050  & 3.96
& 48.22  & 0.517  & $<$0.5   \\
$\ast$MS0011.7$+$0837  &           &00 &11 &45.5  &$+$08 &37 &19.4   &.163
&11.60  & 33.10  & 3.769  & 80.3   \\
 MS0013.4$+$1558  &           &00 &13 &20.9  &$+$15 &58 &17.3   &.083  & 3.59
& 20.83  & 0.616  & $<$1.0   \\
$\ast$MS0015.9$+$1609  &CL0016$+$16  &00 &15 &58.3  &$+$16 &09 &34.0   &.546  &
7.06  & 11.58  &14.639  &  0.8   \\
 MS0026.4$+$0725  &           &00 &26 &26.5  &$+$07 &25 &37.0   &.170  & 4.71
& 13.00  & 1.609  &  2.7   \\
 MS0037.8$+$2917  &A77        &00 &37 &48.1  &$+$29 &16 &53.6   &.069
&18.64  &139.00  & 2.838  &  5.3   \\
 MS0043.3$-$2531  &           &00 &43 &17.5  &$-$25 &32 &06.0   &.112  & 2.26
&  9.16  & 0.493  &  1.8   \\
 MS0102.3$+$3255  &           &01 &02 &21.3  &$+$32 &55 &19.8   &.080  & 7.74
& 47.11  & 1.294  &  1.4   \\
 MS0109.4$+$3910  &           &01 &09 &18.2  &$+$39 &11 &21.6   &.208  & 1.86
&  4.45  & 0.824  &  9.1   \\
 MS0147.8$-$3941  &           &01 &47 &53.2  &$-$39 &42 &01.5   &.373  & 1.47
&  2.66  & 1.581  & $<$0.3   \\
 MS0159.1$+$0330  &A293       &01 &59 &12.0  &$+$03 &30 &28.1   &.165  & 2.96
&  8.38  & 0.977  &260.0   \\
 MS0301.7$+$1516  &           &03 &01 &43.3  &$+$15 &15 &50.6   &.083  & 1.93
& 11.18  & 0.330  & $<$0.8   \\
$\ast$MS0302.5$+$1717  &           &03 &02 &29.4  &$+$17 &16 &47.6   &.425  &
2.15  &  3.75  & 2.879  & $<$1.3 \\
$\ast$MS0302.7$+$1658  &           &03 &02 &43.2  &$+$16 &58 &27.0   &.426  &
3.75  &  6.54  & 5.043  &  3.3 \\
 MS0320.9$-$5322  &           &03 &20 &47.6  &$-$53 &21 &46.5   &.071  & 5.69
& 40.82  & 0.883  &        \\
 MS0353.3$-$7411  &A3186      &03 &53 &15.3  &$-$74 &10 &39.2   &.127  &18.76
& 66.87  & 4.625  &        \\
$\ast$MS0353.6$-$3642  &S400       &03 &53 &38.3  &$-$36 &42 &33.1   &.320  &
6.25  & 11.99  & 5.244  &  1.2   \\
 MS0407.2$-$7123  &           &04 &07 &06.4  &$-$71 &24 &25.3   &.229  & 4.68
& 10.58  & 2.373  &       \\
 MS0418.3$-$3844  &           &04 &18 &20.8  &$-$38 &45 &09.6   &.350  & 1.49
&  2.74  & 1.433  & $<$0.4   \\
 MS0419.0$-$3848  &           &04 &18 &59.6  &$-$38 &49 &01.2   &.225  & 0.78
&  1.78  & 0.385  & $<$0.5   \\
$\ast$MS0433.9$+$0957  &           &04 &33 &58.4  &$+$09 &57 &36.7   &.159
&13.74  & 40.01  & 4.335  & 43.5   \\
$\ast$MS0440.5$+$0204  &           &04 &40 &33.8  &$+$02 &04 &43.7   &.190
&10.19  & 25.91  & 4.007  &  4.9   \\
 MS0450.6$-$5602  &           &04 &50 &36.7  &$-$56 &02 &11.4   &.094  & 2.42
& 12.02  & 0.456  &        \\
$\ast$MS0451.5$+$0250  &A520       &04 &51 &35.7  &$+$02 &50 &34.7   &.202
&16.35  & 39.92  & 6.976  &  5.2 \\
$\ast$MS0451.6$-$0305  &           &04 &51 &40.5  &$-$03 &05 &46.0   &.55  &
9.51  & 15.57  &19.976  & $<$0.8 \\
 MS0508.8$-$4523  &           &05 &08 &50.2  &$-$45 &23 &00.8   &(.20) &16.35
& 40.16  & 6.880  &      \\
 MS0537.1$-$2834  &           &05 &37 &06.8  &$-$28 &34 &40.6   &.254  & 1.02
&  2.17  & 0.599  & $<$1.1   \\
 MS0620.6$-$5239  &           &06 &20 &36.9  &$-$52 &40 &01.7   &.048  & 8.62
&112.00  & 1.107  &1230.0   \\
 MS0623.6$-$5238  &           &06 &23 &40.6  &$-$52 &38 &58.3   &.074  & 2.00
& 13.52  & 0.318  &      \\
 MS0624.3$-$5519  &           &06 &24 &18.6  &$-$55 &19 &17.6   &.118  & 4.76
& 18.28  & 1.091  &       \\
 MS0733.6$+$7003  &A588       &07 &33 &37.8  &$+$70 &03 &39.1   &.117  & 2.14
&  8.29  & 0.487  & $<$0.6   \\
$\ast$MS0735.6$+$7421  &Zw1370     &07 &35 &34.8  &$+$74 &21 &37.4   &.216
&13.10  & 30.64  & 6.119  &  2.3   \\
 MS0810.5$+$7433  &           &08 &10 &36.2  &$+$74 &33 &25.2   &.282  & 2.25
&  4.56  & 1.549  & $<$0.9   \\
$\ast$MS0811.6$+$6301  &           &08 &11 &36.8  &$+$63 &02 &21.3   &.312  &
2.60  &  5.04  & 2.098  & $<$0.8 \\
 MS0821.5$+$0337  &           &08 &21 &33.7  &$+$03 &37 &30.3   &.347  & 1.39
&  2.58  & 1.328  & $<$0.6   \\
$\ast$MS0839.8$+$2938  &Zw1883     &08 &39 &53.3  &$+$29 &38 &16.0   &.194
&13.23  & 33.18  & 5.348  &  5.6   \\
 MS0849.7$-$0521  &           &08 &49 &46.3  &$-$05 &21 &36.5   &.192  & 2.96
&  7.46  & 1.179  &  7.1   \\
 MS0904.5$+$1651  &A744       &09 &04 &33.0  &$+$16 &51 &15.0   &.073  & 5.82
& 40.12  & 0.918  & $<$1.4   \\
$\ast$MS0906.5$+$1110  &A750       &09 &06 &30.1  &$+$11 &10 &39.3   &.180
&15.72  & 41.56  & 5.769  & $<$0.5   \\
 MS0955.7$-$2635  &           &09 &55 &45.2  &$-$26 &35 &56.2   &.145  & 7.19
& 22.63  & 2.039  & $<$0.7   \\
 MS1004.2$+$1238  &           &10 &04 &12.7  &$+$12 &38 &15.9   &.166  & 2.79
&  7.84  & 0.925  &  2.3   \\
$\ast$MS1006.0$+$1202  &Zw2933     &10 &06 &07.3  &$+$12 &02 &20.4   &.221  &
9.99  & 23.05  & 4.819  & $<$0.6   \\
$\ast$MS1008.1$-$1224  &           &10 &08 &05.4  &$-$12 &25 &07.4   &.301  &
5.89  & 11.60  & 4.493  & $<$0.8   \\
 MS1020.7$+$6820  &A981       &10 &20 &36.7  &$+$68 &20 &03.7   &.203  & 2.80
&  6.82  & 1.204  &  1.1   \\
 MS1050.7$+$4946  &           &10 &50 &47.5  &$+$49 &45 &54.4   &.140  &12.47
& 40.52  & 3.405  & 53.8   \\
$\ast$MS1054.4$-$0321  &           &10 &54 &26.9  &$-$03 &21 &25.3   &(.81) &
2.11  &  3.27  & 9.016  &  6.1 \\
 MS1058.7$-$2227  &A1146      &10 &58 &48.0  &$-$22 &27 &46.6   &.141  &12.10
& 39.06  & 3.329  &  2.9   \\
 MS1111.8$-$3754  &           &11 &11 &49.8  &$-$37 &54 &55.8   &.129  &17.23
& 60.34  & 4.325  & $<$0.7   \\
\end{tabular}

\begin{tabular}{rcrrrrrrcrrrr}
 MS1125.3$+$4324  &           &11 &25 &17.9  &$+$43 &24 &09.5   &.181  &20.44
&  5.38  & 0.756  & $<$0.6   \\
 MS1127.7$-$1418  &A1285      &11 &27 &52.3  &$-$14 &18 &19.2   &.105  &13.52
& 58.92  & 2.786  & $<$3.2   \\
$\ast$MS1137.5$+$6625  &           &11 &37 &36.2  &$+$66 &24 &56.9   &(.65) &
1.89  &  3.01  & 5.375  & $<$0.9 \\
$\ast$MS1147.3$+$1103  &           &11 &47 &18.2  &$+$11 &03 &15.9   &.303  &
2.99  &  5.87  & 2.304  &  6.1   \\
 MS1154.1$+$4255  &           &11 &54 &11.6  &$+$42 &54 &49.5   &.174  & 3.57
&  9.69  & 1.258  &  9.8   \\
$\ast$MS1201.5$+$2824  &           &12 &01 &30.4  &$+$28 &23 &47.1   &.167  &
6.05  & 16.94  & 2.025  & $<$1.0   \\
 MS1205.7$-$2921  &           &12 &05 &43.1  &$-$29 &21 &18.0   &.171  & 2.86
&  7.85  & 0.984  &  4.4   \\
$\ast$MS1208.7$+$3928  &           &12 &08 &44.0  &$+$39 &28 &19.0   &.340  &
2.19  &  4.11  & 2.030  &  1.4   \\
$\ast$MS1209.0$+$3917  &           &12 &09 &00.6  &$+$39 &18 &04.6   &(.33)  &
2.81  &  5.33  & 2.493  &  4.6   \\
 MS1219.9$+$7542  &           &12 &19 &56.9  &$+$75 &42 &53.0   &.240  & 2.36
&  5.19  & 1.281  &  1.9   \\
$\ast$MS1224.7$+$2007  &           &12 &24 &42.6  &$+$20 &07 &30.0   &.327  &
5.30  & 10.09  & 4.606  & $<$0.7   \\
$\ast$MS1231.3$+$1542  &           &12 &31 &24.3  &$+$15 &42 &28.4   &.238  &
5.38  & 11.89  & 2.883  & $<$0.6   \\
$\ast$MS1241.5$+$1710  &           &12 &41 &31.6  &$+$17 &10 &06.7   &.312  &
4.23  &  8.20  & 3.411  & $<$0.7   \\
$\ast$MS1244.2$+$7114  &~~Zw5434~~ &12 &44 &09.6  &$+$71 &14 &15.1   &.225  &
7.77  & 17.73  & 3.843  &  5.5   \\
$\ast$MS1253.9$+$0456  &Zw5587     &12 &53 &54.1  &$+$04 &56 &25.7   &.230  &
6.16  & 13.88  & 3.143  & $<$0.6   \\
 MS1305.4$+$2941  &Zw5722     &13 &05 &25.9  &$+$29 &41 &47.0   &.241  & 2.63
&  5.77  & 1.434  &  2.4   \\
 MS1306.7$-$0121  &           &13 &06 &44.8  &$-$01 &21 &23.3   &.088  & 9.50
& 51.13  & 1.698  &  3.1   \\
 MS1308.8$+$3244  &           &13 &08 &50.6  &$+$32 &43 &59.5   &.245  & 3.19
&  6.93  & 1.779  & $<$3.2   \\
 MS1317.0$-$2111  &           &13 &17 &02.2  &$-$21 &11 &38.9   &.164  & 4.63
& 13.14  & 1.515  & $<$1.0   \\
$\ast$MS1333.3$+$1725  &           &13 &33 &21.7  &$+$17 &24 &57.8   &.460  &
3.52  &  6.00  & 5.404  & $<$0.9   \\
 MS1335.2$-$2928  &           &13 &35 &16.2  &$-$29 &29 &19.2   &.189  & 3.30
&  8.43  & 1.289  & 86.0  \\
$\ast$MS1358.4$+$6245  &Zw6429     &13 &58 &20.9  &$+$62 &45 &35.0   &.327
&~~~12.23  & 23.27  &10.624  &  3.8  \\
 MS1401.9$+$0437  &           &14 &01 &56.8  &$+$04 &37 &23.6   &.230  & 1.72
&  3.87  & 0.876  & $<$0.9   \\
 MS1409.9$-$0255  &           &14 &09 &59.8  &$-$02 &55 &07.7   &.221  & 2.16
&  4.98  & 1.042  & $<$1.0   \\
 MS1421.0$+$2955  &           &14 &20 &59.2  &$+$29 &55 &46.0   &.261  & 2.26
&  4.75  & 1.384  & $<$1.0   \\
$\ast$MS1426.4$+$0158  &           &14 &26 &26.7  &$+$01 &58 &36.9   &.320  &
4.42  &  8.47  & 3.707  & $<$1.0   \\
 MS1454.0$+$2233  &Zw7160     &14 &54 &00.3  &$+$22 &33 &15.0   &.108  & 1.79
&  7.55  & 0.377  &  2.8   \\
$\ast$MS1455.0$+$2232  &           &14 &55 &00.5  &$+$22 &32 &34.7   &.259
&~~~26.44  & 55.87  &16.029  &  1.9   \\
$\ast$MS1512.4$+$3647  &           &15 &12 &25.9  &$+$36 &47 &26.7   &.372  &
4.49  &  8.14  & 4.807  &  3.8   \\
 MS1520.1$+$3002  &           &15 &20 &09.4  &$+$30 &03 &16.7   &.117  & 3.46
& 13.40  & 0.787  &  1.4   \\
 MS1522.0$+$3003  &A2069      &15 &22 &03.6  &$+$30 &03 &52.1   &.116  &10.41
& 40.68  & 2.347  & $<$1.2   \\
 MS1531.2$+$3118  &A2092      &15 &31 &14.1  &$+$31 &18 &42.2   &.067  & 2.97
& 23.28  & 0.444  & 12.6   \\
 MS1532.5$+$0130  &           &15 &32 &29.4  &$+$01 &30 &46.3   &.320  & 1.96
&  3.75  & 1.641  & $<$0.9   \\
$\ast$MS1546.8$+$1132  &           &15 &46 &52.0  &$+$11 &32 &25.6   &.226  &
5.90  & 13.44  & 2.937  & $<$1.1   \\
 MS1558.5$+$3321  &A2145      &15 &58 &26.9  &$+$33 &21 &40.7   &.088  & 7.95
& 42.75  & 1.420  &  5.0   \\
$\ast$MS1610.4$+$6616  &           &16 &10 &31.1  &$+$66 &16 &00.0   &(.55) &
2.24  &  3.66  & 4.701  & $<$1.1  \\
 MS1617.1$+$3237  &           &16 &17 &08.8  &$+$32 &37 &52.6   &.274  & 1.80
&  3.69  & 1.185  & $<$0.9   \\
$\ast$MS1618.9$+$2552  &A2177      &16 &18 &56.7  &$+$25 &53 &22.3   &.161  &
7.00  & 20.17  & 2.241  & $<$0.9 \\
$\ast$MS1621.5$+$2640  &           &16 &21 &32.2  &$+$26 &41 &06.4   &.426  &
3.38  &  5.88  & 4.546  &  2.6   \\
 MS1754.9$+$6803  &Zw8303     &17 &54 &50.8  &$+$68 &03 &51.6   &.077  & 9.00
& 57.63  & 1.467  &  2.5   \\
$\ast$MS1910.5$+$6736  &           &19 &10 &29.8  &$+$67 &36 &24.6   &.246  &
7.81  & 16.94  & 4.386  & $<$2.0   \\
$\ast$MS2053.7$-$0449  &           &20 &53 &44.0  &$-$04 &49 &24.7   &.583  &
2.48  &  4.01  & 5.775  & $<$0.8   \\
 MS2124.7$-$2206  &           &21 &24 &39.4  &$-$22 &07 &15.2   &.113  & 5.27
& 21.20  & 1.161  & $<$0.5   \\
$\ast$MS2137.3$-$2353  &           &21 &37 &24.4  &$-$23 &53 &17.6   &.313
&19.28  & 37.33  &15.621  &  1.0   \\
 MS2142.7$+$0330  &           &21 &42 &43.1  &$+$03 &30 &40.0   &.239  & 2.66
&  5.87  & 1.436  & $<$0.9   \\
 MS2215.7$-$0404  &           &22 &15 &41.3  &$-$04 &04 &25.2   &.090  & 6.58
& 34.44  & 1.196  & $<$0.9  \\
 MS2216.0$-$0401  &           &22 &16 &04.7  &$-$04 &01 &51.9   &.090  &10.65
& 55.69  & 1.935  & $<$0.9  \\
$\ast$MS2255.7$+$2039  &Zw8795     &22 &55 &40.6  &$+$20 &39 &04.2   &.288  &
2.87  &  5.76  & 2.041  & $<$0.7   \\
$\ast$MS2301.3$+$1506  &Zw8822     &23 &01 &17.1  &$+$15 &06 &49.8   &.247  &
5.82  & 12.61  & 3.291  &  2.5   \\
 MS2311.2$-$4259  &S1101      &23 &11 &12.1  &$-$43 &00 &00.3   &.058  &93.19
&898.10  &12.963  & 46.4   \\
 MS2316.3$-$4222  &S1111      &23 &16 &21.1  &$-$42 &23 &15.4   &.045  &24.15
&348.80  & 3.031  &540.0   \\
$\ast$MS2318.7$-$2328  &A2580      &23 &18 &47.5  &$-$23 &28 &55.1   &.187
&17.76  & 45.69  & 6.844  & 14.6   \\
 MS2318.9$-$4210  &A3998      &23 &18 &53.8  &$-$42 &10 &14.8   &.089  &21.32
&113.10  & 3.842  &  3.5   \\
 MS2348.0$+$2913  &           &23 &48 &03.5  &$+$29 &13 &01.2   &.095  &14.72
& 72.06  & 2.789  & 14.1   \\
 MS2354.4$-$3502  &A4059      &23 &54 &25.9  &$-$35 &02 &15.8   &.046  & 3.10
& 43.14  & 0.392  &110.0   \\
 MS2356.9$-$3434  &           &23 &56 &53.2  &$-$34 &35 &04.2   &.115  & 1.82
&  7.19  & 0.408  & $<$0.5   \\
& & & & & & & & & & & & \\
\hline
\end{tabular}
}
\normalsize
\end{center}

\clearpage

\begin{center}
\samepage
{\normalsize Notes to Table 1}
\footnotesize
{\sf
\begin{tabular}{rp{5.0in}}
\hline
\multicolumn{1}{c}{Cluster}&\multicolumn{1}{c}{Notes} \\
\hline
 & \\
MS0002.8$+$1556 & Loose cluster with dominant galaxy, extended in
X-rays (contour map in Gioia {\it et~al.} 1990a). \\
$\ast$MS0011.7$+$0837 & The dominant galaxy is a cD radiogalaxy with the
classical double-lobed structure. \\
MS0013.4$+$1558 & This cluster is clumpy in X-rays with some of the
peaks associated with the galaxies. An X-ray contour map has been
presented in Gioia {\it et~al.} (1984). The two bright stars in the
area (a 16th magnitude M star and a 12th magnitude G star) may
contribute part of the emission, however they are both too faint
to be the majority of the total X-ray flux. \\
$\ast$MS0015.9$+$1609 & This is a well known distant cluster (often called
Cl 0016$+$16) with an optical
linear structure. It is extremely rich and it was first discovered by R.
Kron (see Koo 1981). Pointed IPC observations were reported by White,
Silk and Henry (1981) but the presence of this cluster in the EMSS
survey is due to its detection in a separate IPC exposure of
another distant cluster also
discovered by Kron but at a different redshift than this one. The
redshift comes from Dressler and Gunn (1992) who also  give spectra of
many of the cluster members. MS0015.9$+$1609 is one of the brightest
X-ray emitting clusters, and its high redshift has made it a popular
object for cosmological studies.  Koo (1981) first noticed that Cl
0016$+$16 was quite red and was an example of a high-z cluster that did not
exhibit the Butcher-Oemler effect. Spectroscopic observations, however,
revealed a number of ``E+A'' active galaxies indicating a so-called
``active fraction'' consistent with other distant clusters. \\
MS0026.4$+$0725 & Rich and compact cluster. The brightest galaxy has
moderate [O$\thinspace$ II]  $\lambda\thinspace 3727$
emission and radio emission (Stocke {\it et~al.}
1991). Iso-contour X-ray map presented in Gioia {\it et~al.} (1984). \\
MS0037.8$+$2917 & A starburst galaxy at z$=$0.076 is present within
the X-ray contours of this cluster that is identified with Abell 77. Given
the typical X-ray luminosities of starburst galaxies, this galaxy does
not contribute significantly to the X-ray emission observed. \\
MS0043.3$-$2531 & Compact group of galaxies elongated in the EW
direction. An optical study of this cluster by Garilli {\it et~al.} (1992)
shows an irregular poor cluster, much similar to a loose group, with
a blue-galaxy fraction of 26\%. \\
MS0102.3$+$3255 & The bright dominant galaxy has no emission lines
and is radio emitting. \\
MS0109.4$+$3910 & Medium-distant compact and rich cluster. \\
MS0147.8$-$3941 & Distant loose cluster. Its low X-ray luminosity
excludes it from the arc survey sample. \\
MS0159.1$+$0330 & Abell 293 (z=0.165) contains the strong radio galaxy
PKS0159$+$034. X-ray emission consists of three portions (see
Gioia {\it et~al.} 1984 for the X-ray iso-contour map and finding chart of the
area), two of which are associated
with Abell 293 (NE and NW). The southern emission is likely due to
a background QSO at z=1.897 and contributes only a small fraction
to the total X-ray flux. \\
MS0301.7$+$1516 & Poor, irregular nearby cluster. \\
$\ast$MS0302.5$+$1717 & This cluster and MS0302.7$+$1658 below were
detected in an
IPC image whose target was Cl~0303+1706, a distant cluster at a
similar redshift (z=0.4181) discovered optically by Gunn, Hoessel \& Oke
(1986; see also Dressler and Gunn 1992). Fabricant, Bautz \& McClintock (1994)
showed that the three clusters are not gravitationally bound.
Photometric and spectroscopic data for both EMSS clusters
are given in Fabricant, Bautz \& McClintock (1994). \\
$\ast$MS0302.7$+$1658 & Compact cluster, the brightest galaxy shows weak
[O$\thinspace$ II]  $\lambda\thinspace 3727$
emission and radio emission. Mathez {\it et~al.} (1992)
discovered the ``straight'' arc between the two brightest
ellipticals and obtained spectra for 4 galaxies and photometry for 39
galaxies. Giraud (1992) published additional data. In our deep image in
$0''.6$ seeing we easily see the giant lensed arc.
See also previous note. \\
MS0320.9$-$5322 & X-ray emission comes from several condensations
(iso-contour map presented in Gioia {\it et~al.} 1990a). NW blob sits
on dominant galaxy which has weak [O$\thinspace$ II] $\lambda\thinspace 3727$
emission (Stocke {\it et~al.} 1991). \\
$\ast$MS0353.3$-$7411 & Abell 3186, extended X-ray structure (map in Gioia
{\it et~al.} 1990a). Garilli {\it et~al.} (1992) compare the X-ray contours
with the
optical isopleths and point out the similarity of the two morphologies
(both slightly elongated in the NW-SE direction)
suggesting a certain degree of coupling between luminous matter and
hot gas. \\
MS0353.6$-$3642 & The optical CCD image shows a compact cluster with the
three central galaxies embedded in a common envelope. Photometry for
this cluster is given in Garilli {\it et~al.} (1992), who compute
a richness parameter N$_{0.5}$=28 resulting in a
cluster richer than expected in the X-ray-luminosity-N$_{0.5}$ relation
(Bahcall 1980) and similar to MS1358.4$+$6245 (see Luppino {\it et~al.}
1991). They quote  a 29\% fraction of blue galaxies
in this cluster. \\
\end{tabular}

\begin{tabular}{rp{5.0in}}
MS0407.2$-$7123 & This cluster
appears to be very rich in a deep CCD image taken at LCO 40$\arcsec$
telescope (Donahue, private communication). \\
MS0418.3$-$3844 & Poor cluster, no dominant galaxy. \\
MS0419.0$-$3848 & Poor cluster, central galaxy has moderate
[O$\thinspace$ II]  $\lambda\thinspace 3727$. \\
$\ast$MS0433.9$+$0957 & Several bright galaxies, none dominant. Loose, rather
poor cluster. Note this is the lowest redshift object in our 41
cluster subsample.\\
$\ast$MS0440.5$+$0204 & Poor cluster with a very compact,
multiple-nucleus central galaxy
that could be a forming cD galaxy. A circular
gravitational arc structure surrounds the cD envelope. Luppino {\it et~al.}
(1993) compute a lower limit to the mass of the core
of $1.0 \times 10^{14} M_{\solar}$ from
the simple lensing geometry. Donahue, Stocke \& Gioia (1992)
report extended H$_\alpha$ emission
coincident with the core of the cluster, suggesting the presence of
a large cooling flow---an interpretation supported by the presence
of radio emission from the cD. \\
MS0450.6$-$5602 & Compact, poor group. \\
$\ast$MS0451.5$+$0250 & Abell 520. There is not a well defined optical
center in this cluster but several galaxy condensations are visible.
The X-ray source was detected near the edge of the
IPC frame with the observed X-ray emission associated with a tight
clump of galaxies at z=0.202. A more detailed analysis of the X-ray
emission indicates that it may extend an additional 2 arcmin to the
NE. If so, a portion of this source may be due to a poor cluster in that
vicinity at a slightly higher redshift (z=0.22). \\
$\ast$MS0451.6$-$0305 & Distant cluster, given its redshift
(S. Morris, private communication) this is the most X-ray luminous
cluster in the EMSS. The optical image reveals a very rich,
compact cluster with a giant arc to the East of the brightest galaxy.\\
MS0508.8$-$4523 & Optical spectrum has poor S/N. The two central
galaxies seem merged in an SRC J plate. X-ray emission is extended,
iso-contour map given in Gioia {\it et~al.} (1990a). \\
MS0537.1$-$2834 & The NE of two central galaxies has strong
[O$\thinspace$ II]  $\lambda\thinspace 3727$ emission but no radio emission. \\
MS0620.6$-$5239 & Contains the powerful radiosource PKS0620$-$526.
The X-ray structure is extended with several
subcondensations. The X-ray map is presented in Gioia {\it et~al.} (1984). \\
MS0623.6$-$5238 & Field heavily obscured by Canopus (see finding chart
in Gioia {\it et~al.} 1984). The cluster classification is
based upon what little can be seen of this field and is, therefore,
tentative. \\
MS0624.3$-$5519 & The dominant galaxy, ESO 161$-$5, has a red nucleus
and a blue envelope. X-ray iso-contour map presented in Gioia {\it et~al.}
(1990a). \\
MS0733.6$+$7003 & Abell 588, dominant galaxy present, quite rich. \\
$\ast$MS0735.6$+$7421 & X-ray extended, the brightest galaxy has strong
[O$\thinspace$ II]  $\lambda\thinspace 3727$
emission and is radio emitting (Stocke {\it et~al.} 1991). Optical morphology
shows
a moderately poor cluster with a large cD galaxy.\\
$\ast$MS0811.6$+$6301 & Possible background cluster present at z$=$0.49.
CCD image reveals a bright central galaxy with a number of
smaller, nearby companions and a loose grouping of other galaxies in
the background.  \\
$\ast$MS0839.8$+$2938 & The brightest galaxy in this compact cluster has
extremely strong [O$\thinspace$ II] and H$\alpha $ emission with
equivalent widths and line ratios similar to those of cooling flow
clusters. The peak of the X-ray emission is consistent with the
position of the brightest cluster galaxy. A multi-waveband study
of the cluster has been published in Nesci {\it et~al.} (1989). High resolution
ROSAT observations (Nesci {\it et~al.} 1994) show three main peaks of
emission. The emission is centered on the dominant galaxy and both the
ellipticity and the position angle are very similar to those of the galaxy.
The CCD image shows a field whose outer parts contain many spirals and clearly
interacting systems, while the core is compact.\\
MS0849.7$-$0521 & X-ray iso-contour map presented in Gioia {\it et~al.}
(1990a). \\
MS0904.5$+$1651 & This cluster is identified with Abell 744. The
cluster is dominated by a normal giant elliptical
and is extended in X-rays. The iso-contour
map has been published by Kurtz {\it et~al.} (1985) who also present optical
data for the cluster galaxies. The spatial distribution of redshifts
is peculiar with the dispersion within the 150 kpc core radius being
much greater than outside, in this Abell 744 is similar to the nearby
cluster Abell 1060 (Hydra I; Richter \& Huchtmeier 1983). \\
$\ast$MS0906.5$+$1110 & In the IPC image the X-ray emission from
this source largely overlaps with the X-ray emission of the source
MS0906.3$+$1111, identified with a Seyfert galaxy at nearly the
same redshift as the cluster. It is thus difficult to determine
the true extent of the cluster x-ray emission. X-ray iso-contour
map presented in Gioia {\it et~al.} (1990a). The optical image shows a dominant
galaxy with a secondary nucleus to the SW. \\
\end{tabular}

\begin{tabular}{rp{5.0in}}
MS0955.7$-$2635 & The central, non-dominant galaxy has weak [O$\thinspace$ II]
$\lambda\thinspace 3727$ emission. The X-ray emission is extended,
see the iso-contour map in Gioia {\it et~al.} (1990a). \\
MS1004.2$+$1238 & The central dominant cluster galaxy in this loose
cluster has strong [O$\thinspace$ II]  $\lambda\thinspace 3727$
emission and is radio emitting (Stocke {\it et~al.} 1991). \\
$\ast$MS1006.0$+$1202 & In this rich and compact cluster four arcs have been
discovered by \lefevre {\it et~al.} (1994). The field is only $\sim 10'$ from
Arcturus and is heavily contaminated by scattered light, especially in
the blue. An X-ray iso-contour map is presented
in Gioia {\it et~al.} (1990a). \\
$\ast$MS1008.1$-$1224 & Very rich cluster, slightly extended in X-rays, see
iso-contour map in Gioia {\it et~al.} (1990a). The cluster has a circular
distribution
of galaxies surrounding a NS-elongated core.  There is a secondary clump
of galaxies to the N.  Some gravitationally lensed arcs have been reported
by \lefevre {\it et~al.} (1994).  \\
MS1020.7$+$6820 & Abell 981, rich and loose cluster, three
radiosources were detected in this field (see Table 2). None of them is
on the brightest cluster galaxy. \\
MS1050.7$+$4946 & The central dominant galaxy is also radioemitter,
not a rich cluster. \\
$\ast$MS1054.4$-$0321 & Extremely rich, very distant cluster.
A very preliminary redshift of $\sim$ 0.81 is
determined from Ca$\thinspace$ II break only (Stocke, private
communication). If confirmed,
this will be the most distant cluster in the EMSS
catalog. The size
and brightness of the cluster galaxies on the CCD image
are consistent with the assumed redshift. The central galaxy
is a radiosource. \\
MS1058.7$-$2227 & Abell 1146, a rich cluster with a dominant galaxy.
The X-ray emission is extended and shows some clumpiness. It is
centered on the cD galaxy, see X-ray iso-contour map in Gioia {\it et~al.}
(1990a). \\
MS1111.8$-$3754 & Redshift determination comes from nine
galaxies, a dominant bright galaxy is present. X-ray emission is
extended but the source is detected at the edge of the IPC so no
iso-contour map was produced. \\
MS1125.3$+$4324 & The two brightest galaxies in this loose group
have very strong [O$\thinspace$ II] $\lambda\thinspace 3727$
emission. They also have H$\alpha$ emission extended on a scale comparable to
their broad-band image (Donahue, Stocke \& Gioia 1992). This source
was previously classified in Stocke {\it et~al.} (1991) as a cooling flow
galaxy. Due to the presence of a few other galaxies, this source
is now classified as a poor cluster. \\
MS1127.7$-$1418 & The source is identified with Abell 1285. As in the
case of MS0353.3$-$7411 ($=$Abell 3186), previously described, there
is a very tight similarity between the optical and the X-ray
morphologies. They are both elongated in the
NW-SE direction along the axis connecting the 2 brightest cluster
galaxies. The X-ray iso-contour map can be found in Gioia {\it et~al.} (1990a).
\\
$\ast$MS1137.5$+$6625 & This cluster of galaxies has no spectroscopic
redshift determination as yet but it is likely to be at a similar
redshift (z = 0.65) as the target of the IPC observation,
QSO 3C263 located $\sim 20'$ to the S,
based upon CCD photometry (Stocke, private communication). \\
$\ast$MS1147.3$+$1103 & Rich, elongated cluster with an elongated cD. \\
MS1154.1$+$4255 & Rather sparse and very poor group of galaxies. \\
$\ast$MS1201.5$+$2824 & Dominant galaxy present in this moderately rich,
compact cluster. Abell 1455 is about 8$\arcmin $ away. \\
$\ast$MS1208.7$+$3928 &  Compact, rich cluster near a very bright star. \\
$\ast$MS1209.0$+$3917 & This source was originally classified as a ``cooling
flow galaxy'' by Stocke {\it et~al.} (1991) on the basis of the very strong
[O$\thinspace$ II] $\lambda\thinspace 3727$ emission exhibited
by the radiogalaxy 40$\arcsec$ SE (the most southern of three objects)
of the elongated object at the center of the optical CCD image.
This source could actually be a very distant cluster but
additional spectroscopic work is needed. At this time we consider
the classification and ID very uncertain.\\
MS1219.0$+$7542 & Poor cluster with a dominant galaxy showing a weak
[O$\thinspace$ II] $\lambda\thinspace 3727$ emission. \\
$\ast$MS1224.7$+$2007 & The brightest galaxy has very strong [O$\thinspace
$II] $\lambda\thinspace 3727$ emission . This cluster does not appear very
rich in the CCD image, but the images of this cluster
were taken through clouds.\\
$\ast$MS1231.3$+$1542 & Rich, binary cluster elongated in the
NW-SE direction. Both galaxies are of comparable brightness.\\
$\ast$MS1241.5$+$1710 & Rich, loose cluster.  No dominant galaxy.\\
$\ast$MS1244.2$+$7114 & Similar to MS0440.5$+$0204 in being a poor cluster
with a central, multiple-nucleus cD galaxy (up to six nuclei in a common
envelope). The cD galaxy has a 5.5
mJy source and strong [O$\thinspace$ II] $\lambda\thinspace 3727$
emission. The halo of light that envelopes the nuclei is not
symmetrical about any major nuclear condensation nor is it
elliptical in shape (R. Schild private communication). This supports
the idea that this represents an early stage in the formation of
the cD galaxies before they would become relaxed. \\
$\ast$MS1253.9$+$0456 & Very rich cluster with a bright cD elongated
NS along with the cluster.\\
\end{tabular}

\begin{tabular}{rp{5.0in}}
MS1305.4$+$2941 & This cluster lies in SA 57. The redshift for the
brightest galaxy, a member of the Zwicky cluster Zw 1305.4$+$2941 has
been measured by H. Spinrad (see Katgert {\it et~al.} 1983). \\
MS1306.7$-$0121 & Poor cluster with dominant central galaxy.
X-ray iso-contour map presented in Gioia {\it et~al.} (1990a). \\
$\ast$MS1333.3$+$1725 & This source is tentatively identified with a distant,
poor cluster of galaxies. A spectrum of the brightest galaxy is at low S/N and
consistent with z$=$0.465. Additional spectroscopic data are
needed to confirm. \\
$\ast$MS1358.4$+$6245 & An optical and X-ray study of this cluster has been
presented in Luppino {\it et~al.} (1991). They find that the cluster
exhibits the Butcher-Oemler effect with a photometric
blue galaxy fraction of 18\%. Fabricant, McClintock \& Bautz (1991)
present a number of spectra of $\sim$70 galaxies in the core of the field.
They find a specroscopic active galaxy fraction of 17\%. \\
MS1401.9$+$0437  & Loose cluster. No dominant galaxy present. \\
MS1409.9$-$0255  & Poor cluster, no dominant galaxy present. \\
$\ast$MS1426.4$+$0158  & Three galaxies in a common envelope at the center
of the cluster. The bright galaxies visible in the CCD image are at a lower
redshift, maybe two clusters in projection here. Field contains a number
of spirals and possibly interacting systems.\\
MS1454.0$+$2233 & Compact poor cluster. \\
$\ast$MS1455.0$+$2232 & Poor cluster with one and possibly two bright
ellipticals. The galaxy to the E is the brighter and is a radio
emitter. It  has strong $\lambda\thinspace 3727$
[O$\thinspace$ II] emission with an equivalent width of
60\AA ~(Mason {\it et~al.} 1981). An apparent gravitational arc is located
slightly to the NE of this galaxy (see \lefevre {\it et~al.} 1994).
In Donahue (1990) this cluster has the strongest emission line
flux (H$\alpha $ and [N$\thinspace$ II]) of their survey, signifying
a possible cooling flow. \\
$\ast$MS1512.4$+$3647  & Centrally condensed cluster at the X-ray position. The
central galaxy image is highly structured.  There is a bright, blue,
slightly-elongated (EW) object in the NW section of the core.
Filamentary, blue structure is present in the central envelope.\\
MS1522.0$+$3003 & An optical and X-ray study of Abell 2069 has been
published in Gioia {\it et~al.} (1982), where an X-ray iso-contour map is also
presented. The cluster shows multiple condensations in both the X-ray
emission and the galaxy surface density and thus the system as a whole
does not appear relaxed. The close correspondence between the
gas and the galaxy distributions seems to indicate that the galaxies
in this system do map the mass distribution. A possible blue arc was
discovered by Donahue, Stocke \& Gioia (1992) North of the
two central brightest galaxies. However, this cluster is at too low
redshift to enter our 41 cluster sample. \\
MS1531.2$+$3118 & Abell 2092, complex X-ray structure, iso-contour
map presented in Gioia {\it et~al.} (1990a).
Redshift determination by  Hoessel, Gunn \& Thuan (1980). \\
MS1532.5$+$0130 & A Seyfert galaxy at z$=$0.074 is also present in
this field and is given as secondary identification by Stocke {\it et~al.}
(1991). \\
$\ast$MS1546.8$+$1132  &
Rich, elongated cluster with a giant cD.  \\
$\ast$MS1610.4$+$6616 & This source was previously unidentified. The CCD
image shows a very distant and condensed cluster. The spectroscopic
data that we are acquiring are not yet conclusive for the
determination of the redshift, but the cluster is certainly at
z$>$0.5. \\
MS1617.1$+$3237 & cD galaxy present. \\
$\ast$MS1618.9$+$2552 & Also Abell 2177. This rich, nearby cluster has
a central galaxy with an asymmetric surface brightness with a
fan-shaped brightening to the NW.\\
$\ast$MS1621.5$+$2640 & X-ray emission of this distant cluster is resolved
and possibly due to more than one optical counterpart. Only higher
X-ray resolution data may unambiguously describe the structure.
The cluster has been studied by Luppino \& Gioia (1992)
who reported the discovery of a gravitationally lensed arc around the
second brightest galaxy which is also radioemitting. \\
MS1754.9$+$6803  & Two brightest galaxies at the center of the
cluster, slightly extended in X-rays. \\
$\ast$MS1910.5$+$6736  & Moderately poor cluster that has some evidence
of gravitational lensing (see \lefevre {\it et~al.} 1994).
The brightest galaxy has a double nucleus. \\
$\ast$MS2053.7$-$0449  & Until recently, this was the highest redshift
cluster in the EMSS.  Luppino \& Gioia (1992) reported the discovery
of a large arc to the NW of the center.  \\
MS2124.7$-$2206 & The brightest galaxy has a multiple nucleus. A starburst
galaxy is also present in the same cluster but, based upon observed X-ray
luminosities of starbursts, this galaxy is unlikely to contribute
significantly to this source. X-ray iso-contour map is presented in Gioia
{\it et~al.} (1990a). \\
$\ast$MS2137.3$-$2353 & This cluster has been studied by Fort
{\it et~al.} (1992)
and Mellier {\it et~al.} (1993) because of the presence of blue circular
and radial arcs. Their lensing models indicate a very compact
core radius of 50 kpc.\\
\end{tabular}

\begin{tabular}{rp{5.0in}}
MS2142.7$+$0330 & cD galaxy present. \\
MS2215.7$-$0404 & and MS2216.0$-$0401 are two separate sources
probably parts of the extended X-ray emission from a single cluster. \\
$\ast$MS2255.7$+$2039 & The M star to the West of the central galaxy
could contribute to the X-ray emission of this cluster. \\
$\ast$MS2301.3$+$1506 & Rich cluster with a dominant cD. \\
MS2311.2$-$4259  & Sersic 159-03. The nuclear region of the cD, ESO 291-G9,
has been classified spectroscopically as LINER by Maia {\it et~al.} (1987). \\
MS2316.3$-$4222 & Southern Abell cluster with very bright cD galaxy.
It contains the powerful radiosource PKS2316$-$423. \\
$\ast$MS2318.7$-$2328 & Also Abell 2580.  Bright, nearby cD galaxy in
a rich cluster. Contains a possible lensed arc
(see \lefevre {\it et~al.} 1994). \\
MS2318.9$-$4210 & Abell 3998. A cD galaxy is present and the X-ray
structure is extended (see iso-contour map in Gioia {\it et~al.} 1990a). \\
MS2348.0$+$2913 & Rich cluster, no dominant galaxy. \\
MS2354.5$-$3502 & Very extended X-ray emission centered on the bright
cD galaxy (ESO 349-G10 possible LINER in Maia {\it et~al.} 1987) which is
also PKS2354$-$350. X-ray iso-contour
map presented in Gioia {\it et~al.} (1987). \\
& \\
\hline
\end{tabular}
}
\normalsize
\end{center}

\clearpage

\begin{center}
\samepage
{\normalsize Table 2.  Radio sources associated with EMSS clusters }
\footnotesize
{\sf
\begin{tabular}{rrrrrrrrp{2.0in}}
\hline\hline
\multicolumn{1}{c}{(1)} & \multicolumn{6}{c}{(2)} &
\multicolumn{1}{c}{(3)} & \multicolumn{1}{c}{(4)}\\
& & & & & & &
\multicolumn{1}{c}{Radio} & \\
& \multicolumn{6}{c}{Radio Coordinates (1950.0)} &
\multicolumn{1}{c}{6\thinspace cm flux} & \\
\multicolumn{1}{c}{Name} & \multicolumn{3}{c}{R.A.} & \multicolumn{3}{c}{Dec.}
&
\multicolumn{1}{c}{(mJy)} & \multicolumn{1}{c}{Comments} \\
\hline
 & & & & & & & & \\
$\ast$MS0011.7$+$0837    &00 &11 &45.4 &$+$08 &37 &22.0  &  64.6   &double
radio source associated with the central galaxy \\
                    &00 &11 &46.5 &$+$08 &37 &15.0  &  15.7   &second
component of double\\
$\ast$MS0015.9$+$1609    &00 &15 &59.5 &$+$16 &10 &03.5  &   0.8   &on galaxy
$\sim 35\arcsec$ to the NE\\
 MS0026.4$+$0725    &00 &26 &26.5 &$+$07 &25 &37.0  &   2.7   &on brightest
galaxy\\
 MS0037.8$+$2917    &00 &37 &48.1 &$+$29 &16 &53.6  &   5.3   &on brightest
galaxy\\
 MS0043.3$-$2531    &00 &43 &17.0 &$-$25 &31 &40.5  &   1.8   &not on brightest
galaxy\\
                    &00 &43 &15.4 &$-$25 &31 &58.0  &   8.4   &double
radio source associated with an AGN (z$=$0.86) to the West\\
                    &00 &43 &14.6 &$-$25 &32 &01.5  &   3.9   &second
component of double\\
 MS0102.3$+$3255    &01 &02 &21.3 &$+$32 &55 &19.8  &   1.4   &on brightest
galaxy\\
 MS0109.4$+$3910    &01 &09 &17.8 &$+$39 &11 &24.6  &   9.1   &on brightest
galaxy\\
 MS0159.1$+$0330    &01 &59 &15.6 &$+$03 &28 &42.6  & 260.0   &PKS0159$+$034,
radio position is $2.\arcmin\thinspace 3 $ SE of X$-$ray position\\
$\ast$MS0302.7$+$1658    &03 &02 &43.2 &$+$16 &58 &27.0  &   3.3   &on
brightest galaxy\\
$\ast$MS0353.6$-$3642    &03 &53 &40.6 &$-$36 &42 &23.1  &   1.2   &on NE
galaxy $\sim 30\arcsec$ away\\
                    &03 &53 &30.1 &$-$36 &43 &04.6  &   3.6 &on galaxy $\sim
1.\arcmin\thinspace 73$ to the East\\
$\ast$MS0433.9$+$0957    &04 &33 &58.4 &$+$09 &57 &36.7  &   2.0   &on bright
galaxy North\\
                    &04 &33 &59.5 &$+$09 &57 &07.9  &  43.5   &on bright galaxy
North$-$East\\
$\ast$MS0440.5$+$0204    &04 &40 &33.8 &$+$02 &04 &43.7  &   4.9   &on bright
galaxy part of multiple nucleus, Luppino {\it et~al.} 1993\\
                    &04 &40 &33.5 &$+$02 &04 &28.7  &   1.8   &on bright galaxy
to the South\\
$\ast$MS0451.5$+$0250    &04 &51 &24.2 &$+$02 &52 &55.8  &   6.1   &on spiral
galaxy to the NW at the edge of CCD \\
                    &04 &51 &39.8 &$+$02 &50 &43.8  &   5.2   &on
galaxy to the NE close to the center declination\\
                    &04 &51 &43.5 &$+$02 &50 &09.8  &   3.6   &on galaxy SE\\
 MS0620.6$-$5239    &06 &20 &37.3 &$-$52 &40 &01.0  &1230.0   &PKS0620$-$526\\
$\ast$MS0735.6$+$7421    &07 &35 &34.8 &$+$74 &21 &37.4  &   2.3   &on
brightest galaxy\\
$\ast$MS0839.8$+$2938    &08 &39 &53.3 &$+$29 &38 &16.0  &   5.6   &on
brightest galaxy\\
 MS0849.7$-$0521    &08 &49 &47.3 &$-$05 &22 &10.2  &   7.1   &not on brightest
galaxy\\
 MS1004.2$+$1238    &10 &04 &12.7 &$+$12 &38 &15.9  &   2.3   &on brightest
galaxy\\
                    &10 &04 &10.3 &$+$12 &41 &12.9  &   3.1   &on galaxy to the
NW\\
 MS1020.7$+$6820    &10 &20 &46.6 &$+$68 &21 &10.6  &   1.1   &not on brightest
galaxy\\
                    &10 &20 &46.1 &$+$68 &20 &57.6  &   0.9   &not on brightest
galaxy\\
                    &10 &20 &52.2 &$+$68 &19 &09.6  &   4.3   &not on brightest
galaxy\\
 MS1050.7$+$4946    &10 &50 &47.5 &$+$49 &45 &54.4  &  53.8   &on brightest
galaxy\\
$\ast$MS1054.4$-$0321    &10 &54 &26.9 &$-$03 &21 &25.3  &   6.1   &on
brightest galaxy\\
 MS1058.7$-$2227    &10 &58 &52.4 &$-$22 &27 &12.4  &   2.9   &on galaxy to the
East\\
                    &10 &58 &39.0 &$-$22 &26 &51.4  &   2.3   &on object far
away from center\\
$\ast$MS1147.3$+$1103    &11 &47 &21.4 &$+$11 &04 &13.3  &   6.1   &on galaxy
to the NE\\
 MS1154.1$+$4255    &11 &54 &11.6 &$+$42 &54 &49.5  &   9.8   &on brightest
galaxy\\
 MS1205.7$-$2921    &12 &05 &43.1 &$-$29 &21 &18.0  &   4.4   &on brightest
galaxy\\
                    &12 &05 &37.5 &$-$29 &23 &01.5  &   1.8   &no object
visible on POSS\\
$\ast$MS1208.7$+$3928    &12 &08 &44.0 &$+$39 &28 &19.0  &   1.4   &on
brightest galaxy\\
$\ast$MS1209.0$+$3917    &12 &09 &02.0 &$+$39 &17 &34.6  &   4.6   &on
galaxy SE of optical CCD center\\
 MS1219.9$+$7542    &12 &19 &56.9 &$+$75 &42 &53.0  &   1.9   &on brightest
galaxy\\
$\ast$MS1244.2$+$7114    &12 &44 &09.6 &$+$71 &14 &15.1  &   5.5   &on
brightest galaxy\\
 MS1305.4$+$2941    &13 &05 &25.9 &$+$29 &41 &47.0  &   2.4   &21 cm flux
(Katgert {\it et~al.} 1983), on brightest galaxy \\
 MS1306.7$-$0121    &13 &06 &44.8 &$-$01 &21 &23.3  &   3.1   &on brightest
galaxy\\
 MS1335.2$-$2928    &13 &35 &16.2 &$-$29 &29 &19.2  &  ~~~~~86.0   &on
brightest galaxy\\
\end{tabular}

\begin{tabular}{rrrrrrrrp{2.0in}}
$\ast$MS1358.4$+$6245    &13 &58 &20.9 &$+$62 &45 &35.0  &   3.8   &on
brightest galaxy\\
 MS1454.0$+$2233    &14 &54 &00.3 &$+$22 &33 &15.0  &   2.8   &on brightest
galaxy\\
$\ast$MS1455.0$+$2232    &14 &55 &00.5 &$+$22 &32 &34.7  &   1.9   &on
brightest galaxy\\
$\ast$MS1512.4$+$3647    &15 &12 &25.9 &$+$36 &47 &26.7  &   3.8   &on
brightest galaxy  \\
 MS1520.1$+$3002    &15 &20 &09.4 &$+$30 &03 &16.7  &   1.4   &not on brightest
galaxy\\
 MS1531.2$+$3118    &15 &31 &13.7 &$+$31 &17 &46.0  &  12.6   &on second
brightest galaxy\\
 MS1558.5$+$3321    &15 &58 &26.9 &$+$33 &21 &40.7  &   5.0   &on brightest
galaxy\\
$\ast$MS1621.5$+$2640    &16 &21 &31.9 &$+$26 &41 &19.9  &   2.6   &on
second brightest galaxy, Luppino \& Gioia (1992)\\
 MS1754.9$+$6803    &17 &54 &50.8 &$+$68 &03 &51.6  &   2.5   &on East
of two central galaxies\\
$\ast$MS2137.3$-$2353    &21 &37 &24.4 &$-$23 &53 &17.6  &   1.0   &on
brightest galaxy (Stocke private communication)\\
$\ast$MS2301.3$+$1506    &23 &01 &17.1 &$+$15 &06 &49.8  &   2.5   &on
brightest galaxy\\
 MS2311.2$-$4259    &23 &11 &12.1 &$-$43 &00 &00.3  &  46.4   &on brightest
galaxy\\
 MS2316.3$-$4222    &23 &16 &22.4 &$-$42 &23 &40.0  & ~~~~540.0   &on
brightest galaxy, PKS2316$-$423\\
$\ast$MS2318.7$-$2328    &23 &18 &47.5 &$-$23 &28 &55.1  &  14.6   &on
brightest galaxy\\
                    &23 &18 &49.9 &$-$23 &29 &15.1  &   1.6   &on second
brightest galaxy\\
 MS2318.9$-$4210    &23 &18 &53.8 &$-$42 &10 &14.8  &   3.5   &on brightest
galaxy\\
 MS2348.0$+$2913    &23 &48 &04.2 &$+$29 &12 &49.4  &   9.3   &double radio
source on brightest galaxy\\
                    &23 &48 &03.1 &$+$29 &13 &03.4  &   4.8   &second
component of double\\
 MS2354.4$-$3502    &23 &54 &26.6 &$-$35 &02 &26.0  & 110.0   &PKS2354$-$350\\
 & & & & & & & & \\
\hline
\end{tabular}
}
\normalsize
\end{center}
\clearpage
{\samepage\small\twocolumn

}

\end{document}